\documentclass[journal, 10]{IEEEtran}
\usepackage{amsmath,amsfonts}
\usepackage{multirow}
\usepackage{algorithm}
\usepackage{algpseudocode}
\usepackage{listings,multicol}
\usepackage{array}
\usepackage{textcomp}
\usepackage{stfloats}
\usepackage{balance}
\usepackage{url}
\usepackage{amssymb}
\usepackage{pifont}
\usepackage{verbatim}
\usepackage{graphicx}
\usepackage{cite}
\usepackage[table]{xcolor}  
\definecolor{mycolor}{rgb}{1,0.0,0.5}
\newcommand{\remark}[1]{}

\usepackage{subcaption}
\usepackage{url}

\usepackage{breakurl}
\usepackage{minted}
\usepackage{flushend}
\usepackage{tikz}

\usepackage[breaklinks]{hyperref}
\usepackage{svg}
\begin{document}

\title{A Dynamic UPF Fault Recovery Mechanism for Enhanced Resilience in 5G Core Networks}

\author{
Shirin Behnaminia$^{1*}$, Pardis Yavari$^{1*}$, Zeinab Zali$^{1}$, Mohammad Reza Heidarpour$^{1}$\\
$^{1}$Department of Electrical and Computer Engineering, Isfahan University of Technology, Isfahan, Iran
}



\maketitle

\def\thefootnote{*}\footnotetext{These authors contributed equally to this work.}

\begin{abstract} 
Ensuring fault tolerance in the User Plane Function is essential to maintain service continuity in 5G networks, particularly as latency-sensitive applications become more prevalent. This paper presents a novel mechanism for dynamic UPF failure detection and recovery within an OpenAirInterface-based 5G core environment. The approach introduces an application-layer session restoration mechanism to ensure service continuity and minimize disruptions during UPF failures\footnote{The source code is available at \url{https://github.com/ION-Lab-Organization/smf}.}. Experimental results demonstrate that the proposed mechanism reduces UPF downtime while maintaining low packet loss and enabling rapid service recovery. In tests with TCP and UDP traffic, connectivity was restored within a few seconds, minimizing service degradation. Furthermore, our failure-aware UPF selection mechanism improves long-term system resilience by considering recent failure history. The results highlight the effectiveness of our lightweight framework in improving the overall reliability of 5G networks, making it a valuable contribution to network robustness.
\end{abstract}


\section{Introduction}
Communication networks have evolved significantly and become more complex over time \cite{ref21}. As part of this progress, 5G has been developed to support diverse services, such as critical communications, industrial IoT, and high-speed internet for mobile users \cite{ref22}.

As a result, ensuring continuous and reliable user connectivity in 5G networks has become more critical than ever. Moreover, adopting Virtual Network Functions (VNFs) has made networks more flexible and scalable, but this shift has increased reliance on software. VNFs offer great advantages such as increasing network scalability and improving resource management \cite{ref23}. However, their operation on general-purpose servers lacks the inherent reliability of traditional dedicated network appliances, which raises concerns about potential failures. A failure in a VNF can interrupt the services that depend on it, making effective fault tolerance essential for maintaining continuous network operation. This is particularly important in 5G networks, where VNFs play a vital role in maintaining overall network functionality. Additionally, 5G networks incorporate network slicing, which allows multiple logical networks to operate over shared physical infrastructure. While this enables customized Quality of Service (QoS) for different applications, it also increases network complexity, making traditional fault management techniques insufficient \cite{ref21}.

A key challenge in this regard is maintaining the fault tolerance of core network components like the User Plane Function (UPF), which plays a unique and essential role in the data plane by directly handling user data traffic. The UPF is also responsible for managing packet forwarding, enforcing QoS policies, and ensuring seamless interconnection between users and external networks such as the internet or private data centers. Because it operates at the heart of the data path, any failure in the UPF can lead to major service disruptions, directly affecting user experience and network performance. This module faces high traffic loads, especially in 5G networks that provide high throughput, low latency, and massive connectivity. As the number of end users and demand for services like video streaming and gaming increases, the UPF’s load also rises. It directly impacts user experience and network performance, especially in a dynamic environment where real-time data processing is essential.

Low latency, a critical feature of 5G networks, relies heavily on the UPF’s ability to process and forward data with minimal delay. This is especially important for latency-sensitive applications such as cloud gaming and augmented or virtual reality (AR/VR), where even slight delays can disrupt user experience and responsiveness. Given the reliance of these services on the UPF, any disruption can lead to severe service interruptions and potentially critical failures, resulting in user dissatisfaction and degraded service quality.
Maintaining session continuity during UPF failures remains a major challenge in cloud-native 5G core deployments. While infrastructure-level tools such as Kubernetes and Docker offer effective mechanisms for scalability and fault recovery, such as automatic restarts, they inherently treat network functions like the UPF as stateless components. As a result, they do not preserve in-memory session context, which is essential for maintaining uninterrupted user sessions. When a UPF is restarted, for instance, it is reinitialized as a completely new instance. This results in the loss of session-specific data, such as PDU session contexts. Consequently, disruptions occur for end users who were already connected to the network, leading to issues such as dropped connections or degraded service, as the re-established UPF lacks the ability to restore the previous state.

Kubernetes provides stateful facilities, such as StatefulSets, designed to manage applications that require stable network identities and persistent storage across restarts. However, these mechanisms are primarily focused on retaining data on persistent volumes and ensuring ordered deployment and scaling. 

In addition, Kubernetes and Docker offer a feature known as Service, which ensures stable access to a group of pods or containers, even as they are replaced or rescheduled. While Services facilitate stable networking and load balancing, they are primarily designed for stateless applications. As such, they do not ensure the continuity of active connections during failures. 

Several recent studies, including [4] and [5], have demonstrated the deployment of 5G core components on Kubernetes to enhance scalability and reliability. However, these efforts largely focus on infrastructure-level concerns and do not address the critical need for preserving session state during failures. This highlights the necessity for application-layer solutions that can maintain session continuity and minimize service disruptions.

To address this limitation, we need to incorporate session restoration logic at the application layer. By explicitly transferring the session context from the Session Management Function (SMF) to the failed UPF during recovery, we ensure continuity in data flows and mitigate user-perceived service disruptions.

This paper introduces a proactive failure detection and recovery mechanism for the UPF to overcome these limitations. Our primary contribution is as follows:
\begin{enumerate}
    \item Leveraging standard procedures, we designed a lightweight mechanism to restore lost context and ensure service continuity efficiently.
    \item Through rigorous testing and performance analysis, we demonstrate the proposed mechanism’s effectiveness in minimizing downtime and ensuring robust service continuity, providing insights for future enhancements in fault-tolerant 5G networks.
 \item Additionally, this work contributes to the OpenAirInterface project by providing a detailed explanation of the codebase and by open-sourcing our implementation. This clarity facilitates the development of new features, empowering contributors to improve UPF functionality and network fault management. 
\end{enumerate} 
The rest of the paper is organized as follows. We begin by reviewing existing studies on fault tolerance in 5G networks to highlight current limitations. Next, we provide background on 5G architecture and core components, with a focus on the SMF, UPF, and PDU session procedures. We then introduce our proposed method for UPF fault recovery, including design considerations such as recovery scenarios, session context restoration, and failure-aware UPF selection. This is followed by a detailed explanation of the OpenAirInterface SMF codebase, where our modifications were implemented. We describe the implementation process, outline the experimental setup, and evaluate the performance of our solution. Finally, we conclude with a summary and suggestions for future work.

\section{Related Work}\label{Sec3}
In related existing works, fault tolerance in 5G networks has been approached from various perspectives, including Artificial Intelligence and system-level frameworks. In the context of fault tolerance in 5G networks, key concepts include state management, storage functions, self-organizing networks (SON), and self-healing mechanisms. These methods are crucial for ensuring resilience and recovery in the event of failures. The following related studies explore these approaches in detail.

Deep learning techniques have enhanced fault tolerance strategies by enabling more accurate failure detection and adding predictive capabilities that help anticipate potential disruptions \cite{ref20}. In \cite{ref11}, the authors present a fault diagnosis method for 5G networks that uses real-world data to avoid overly idealized simulation results. They apply a combination of Generative Adversarial Networks, Naive Bayes models, and graph-based models to enhance dataset quality and identify fault patterns in the network. Similarly, \cite{ref12} proposes a time series-based approach for predicting failures within the 5G core. By collecting traffic load data from the Open5GCore platform and applying machine learning algorithms for forecasting, their method aims to detect anomalies before failures occur. While both works demonstrate strong capabilities in fault detection and prediction, they do not address how to manage recovery after faults—an aspect this paper directly addresses.

Another relevant study \cite{ref13} focuses on Self-Healing mechanisms in 5G networks using a context-aware proactive approach powered by machine learning. Their framework consists of detection and compensation processes, where K-means clustering is used to classify the system into different faulty states. The compensation phase leverages a Markov Decision Process model to determine optimal recovery actions before an outage occurs. The study demonstrates that their optimized approach outperforms random action selection. The algorithm is tested in a network simulator that models small cells in different states. While their work contributes to intelligent SONs, it primarily investigates theoretical modeling rather than real-world deployment.

Beyond AI-driven methods, a complementary line of research focuses on systematic and architecture-level approaches. These methods emphasize structural design choices—such as state management, redundancy, and failover mechanisms—to enhance the resilience of 5G core networks. The following review explores key studies that adopt structured methodologies to improve fault tolerance in 5G systems.

Several studies have explored the use of the Unstructured Data Storage Function (UDSF) as a means to improve state management and fault tolerance in 5G core networks. UDSF provides a centralized repository for storing application state and user equipment (UE) context, enabling the decoupling of state information from individual network functions.

A closely related study \cite{ref14} demonstrates this approach by starting with a stateful implementation of a production-grade 5G core. The authors then modify the network functions to offload application state and UE context to an external UDSF, effectively transforming the functions into stateless components. This approach enhances scalability and failover by leveraging Kubernetes’ capabilities. The work includes a comparison of different design choices, examining the trade-offs between stateful and stateless architectures. However, it has some limitations. It focuses only on AMF and AUSF, leaving the data plane unaddressed. Also, the transition to a fully stateless core ensures 100\% successful registrations but at the cost of higher signaling overhead. Additionally, the implementation is not open-source, restricting its accessibility for further research and development.

ML-SLD \cite{ref17} introduces a cloud-native 5G core network implementation based on a Message-Level Stateless principle, where individual NAS and NGAP messages serve as the unit of processing. Unlike traditional architectures that maintain session states, ML-SLD retrieves all necessary context from remote storage (UDSF), making core network functions pure protocol-processing entities. Additionally, it proposes a non-3GPP-defined Network Function (NF) that acts as middleware between the RAN and AMF, handling messages over SCTP and HTTP. While this work focused on the control plane, it did not address the data plane, unlike our approach.

PP5GS \cite{ref19} introduces a procedure-based, stateless 5G core by implementing a 3GPP-compliant UDSF to store and retrieve states remotely. This design improves CPU utilization, reduces procedure completion time, and enhances edge offloading. However, it achieves these benefits by tightly integrating all functionalities within a single NF, which reduces signaling overhead but also sacrifices modularity and reusability. This centralized approach creates scalability bottlenecks and introduces a single point of failure under high network loads.

Apart from UDSF-based approaches, other works have explored different strategies to improve fault tolerance in 5G networks. Authors in \cite{ref15} propose a fault-tolerant 5G authentication mechanism called FT5AS, which introduces a Mediator to track and manage the authentication steps of UEs. The Mediator maintains a status table recording the authentication state of each UE, ensuring that if an AUSF instance fails, another instance can seamlessly take over its authentication tasks using information from this table. The system relies on VMware, which enables automatic restart and recovery of VMs in case of failure. However, their implementation is limited to a Mininet-based simulation rather than a real-world deployment. Moreover, it focuses solely on the control plane, without addressing data plane resilience.

Another related work \cite{ref16} proposes a 1+1 path protection mechanism for 5G networks using Packet Replication, Elimination, and Ordering (PREOF). This design places two intermediate nodes at opposite ends of a communication tunnel. The first node duplicates user data packets and sends them simultaneously to two UPFs. The second node receives packets from both UPFs, discards duplicates, and forwards only one copy onward. This redundancy ensures uninterrupted data flow if one UPF fails. The method delegates packet reordering to external servers to reduce latency and packet loss. Although the approach is theoretically innovative and thoroughly outlined, its practical implementation and performance evaluation are left for future work.

A further contribution in this area is found in \cite{ref18}, which focuses on the control plane, employing a primary-backup replication scheme with asynchronous updates. When a Control Plane Function (CPF) fails, the Control Traffic Aggregator (CTA) replays logged control messages to restore state, accelerating recovery by limiting interactions to signaling between CPF and CTA. However, this approach sacrifices consistency due to asynchronous state updates, which may not capture recent changes if a CPF fails before syncing with the CTA. Additionally, if the CTA fails, all logged states are lost, requiring User Equipment to reattach to the network, further highlighting the trade-off between speed and consistency. 

While prior research has explored fault tolerance in 5G networks through AI-driven prediction and architectural solutions, achieving efficient recovery with low overhead remains a challenge. Our approach builds on these efforts by specifically addressing failures in the UPF. Unlike existing methods that primarily focus on the control plane or rely on complex redundancy mechanisms, our solution enables the UPF to restore its state seamlessly after a restart, with the SMF providing the necessary context. This ensures continuity of user sessions with low signaling overhead, offering a lightweight and effective fault-tolerance solution for the data plane.

\section{Background}\label{Sec2}
This section provides an overview of the essential 5G components relevant to our proposed fault-tolerance mechanism. We briefly introduce the 5G architecture and highlight the roles of the SMF and the UPF. We then explain the Packet Data Unit (PDU) session establishment process, emphasizing how session state is distributed across the network.

\subsection{5G Network Overview }

5G, the fifth generation of cellular technology, enhances speed, reduces latency, and improves service flexibility by supporting features like network slicing, scalability, security, efficient content delivery, and interworking with earlier technologies. In 5G, the system architecture comprises key elements familiar from previous generations, such as User Equipment (UE), the Radio Access Network (NG-RAN), and the Core Network (5GC), which work together to provide end-to-end connectivity.

The 5G Core is structured around a Service-Based Architecture (SBA) that departs from traditional network entities. Instead of rigid, monolithic functions, 5GC introduces modular Network Functions (NFs), which interact with each other through standardized service-based interfaces. This architecture enhances modularity and reusability, as any NF can offer its services to authorized consumers within the network. 

A foundational principle of the 5GC architecture is the separation of the Control Plane (CP) and the User Plane (UP). CP functions such as the AMF and SMF handle signaling and session control, while the UPF manages user data forwarding. 

\begin{figure}[!htbp]
	\centering{\includegraphics[width=0.48\textwidth]{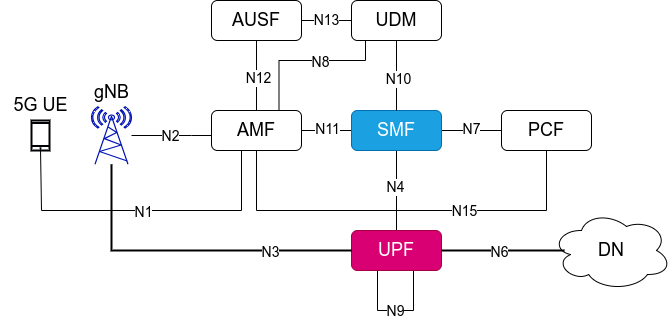}}
	\caption{5G Standalone Service-Based Architecture Showing Key Network Functions and their Standard Interfaces\label{fig:fig1}}
\end{figure}

Within a PDU session, the SMF maintains and manages the session context, including metadata such as QoS parameters, traffic rules, and network slice information. The UPF, on the other hand, retains the actual forwarding state and performs real-time processing of user-plane traffic based on instructions from the SMF, communicated via the N4 interface using the PFCP protocol as shown in Figure. \ref{fig:fig1} \cite{ref4}.

This tight coordination between the SMF and UPF is essential for session continuity, performance, and policy enforcement. Therefore, our focus on these two Network Functions is motivated by their joint responsibility in managing PDU sessions: the SMF governs the control signaling, while the UPF executes the corresponding data-plane behavior. Any failure or misalignment between them can lead to session termination, service disruption, or broader network instability.


\subsection{PDU Session Establishment}
Establishing a data connection in 5G requires aligning a series of context elements across multiple network functions. These contexts define how traffic should be identified, routed, and treated throughout the network. This section explores the types of context exchanged and how they together result in the formation of a PDU Session.

\begin{figure*}[t]	
\centering{\includesvg[width=0.8\textwidth]{figs/pdu-session-procedure}}
	\caption{This figure depicts the signaling procedure for session establishment. It showcases the key data transmitted and includes the context for each module at the conclusion of the process.\label{fig:fig2}}
\end{figure*}

\subsubsection{Context Exchange for Session Setup}
Several entities—including the UE, AMF, SMF, UPF, and gNB, coordinate before any user data can be transmitted. In this section we describe the nature and sequencing of the information exchanged among the AMF, SMF, gNB, and UPF — as illustrated in Fig. ~\ref{fig:fig2}.
The process begins with the AMF collecting the UE's identity and key information needed for session setup. This includes:
\begin{itemize}
    \item UE and Slice Identification: The AMF collects the Subscriber Permanent Identifier (SUPI), Data Network Name (DNN), and the Single-Network Slice Selection Assistance Information (S-NSSAI) from the UE and forwards them to the SMF as input for session creation, enabling proper network slice selection.
\end{itemize}

The SMF, as the session orchestrator, uses this input to construct a full session context, which includes:

\begin{itemize}
    \item Session Parameters:
    \begin{itemize}
        \item[1] PDU Session ID
        \item[2] Session type (e.g., IPv4, IPv6, Ethernet)
        \item[3] IP address allocation for the UE
        \item[4] Policy and charging rules
    \end{itemize}

    \item QoS Configuration:
         A set of QoS Flow Identifiers (QFIs), each linked to a specific traffic class and associated with a QoS profile (e.g., latency, priority, reliability). These QFIs are later mapped to Data Radio Bearers (DRBs) on the RAN side.
\end{itemize}

The SMF then selects a UPF and coordinates with it to establish the uplink tunnel endpoint. This involves:
\begin{itemize}
    \item Tunnel Information (Uplink):
        The UPF allocates an uplink F-TEID and IP address. These are returned to the SMF and propagated toward the gNB (via AMF).
\end{itemize}

Next, the SMF prepares the initial session context, including:
\begin{itemize}
    \item Traffic Handling Rules:
        \begin{itemize}
            \item[1] Packet Detection Rules (PDRs): Define how incoming packets should be identified and classified.
            \item[2] Forwarding Action Rules (FARs): Specify the actions to be applied to matched packets, such as forwarding, buffering, or dropping.
        \end{itemize}
\end{itemize}
These rules are installed at the UPF and define its behavior for this session \cite{ref7}.

At this point, the AMF delivers this context to the gNB, which proceeds with:
\begin{itemize}
    \item RAN Resource Configuration:
        \begin{itemize}
            \item[1] The gNB allocates DRBs based on QoS requirements.
            \item[2] It creates a downlink tunnel endpoint and returns its F-TEID and IP address.
        \end{itemize}
\end{itemize}
    
Finally, the SMF completes the context exchange by updating the UPF with the gNB’s downlink tunnel info:
\begin{itemize}
    \item Tunnel Information (Downlink): The Downlink F-TEID and gNB IP are sent to the UPF. This enables the UPF to forward downlink packets to the gNB, completing the tunnel.

\end{itemize}
At this point, all tunnel and QoS contexts are aligned across the network. The session is considered established, and data can begin flowing through the PDU session tunnel \cite{ref5}.

\subsubsection{PDU Session as a Logical Tunnel}
All of the above context — UE identity, session parameters, QoS flows, tunnel endpoints, and forwarding rules — collectively form the PDU Session. A PDU itself is a block of data exchanged between systems, encapsulated with protocol-specific information as it moves through network layers. Essentially, a PDU Session acts as a tunnel for data transmission, facilitating the flow of information between the UE and the Data Network while passing through both the gNB and the UPF, which act as the tunnel gateways \cite{ref1}.

As illustrated in Fig. \ref{fig:fig3}, a PDU session is composed of one or more QoS Flows, each identified by a QFI and designated for specific types of traffic. Each QoS Flow traverses the PDU session tunnel, which consists of two components:
\begin{itemize}
    \item[1] UP Tunnel: Between UPF and gNB for transmitting user plane data in the IP protocol.
    
    \item[2] Data Radio Bearer (DRB): Between gNB and UE for transmitting user plane data in LTE protocol.
\end{itemize}

\begin{figure}[t]
	\centering{\includesvg[width=0.35\textwidth]{figs/pdu-session}}
	\caption{PDU Session logical structure in 
 5G\label{fig:fig3}}
\end{figure}

\section{Proposed Method}\label{Sec4}

Cloud-native fault tolerance mechanisms are primarily designed for stateless applications, making traditional recovery approaches less effective for stateful components like the UPF. To address this, we implement context transfer at the application layer, allowing the UPF to restore its state without relying on complex redundancy schemes. Our approach assumes a basic infrastructure where each module has a static IP and an automatic restart mechanism, ensuring a lightweight yet reliable failure recovery process.

\subsection{Fault Tolerance Procedure}

The fault tolerance procedure is initiated when the SMF detects a UPF startup event, characterized by two key conditions:

\begin{itemize}
    \item The SMF detects that a UPF is in the process of starting up.
    \item The SMF keeps a list of registered UPFs with their IP addresses and profiles, and by matching the IP, it identifies this UPF as already registered.
\end{itemize}

This scenario signals that a failure has occurred prior to the restart, prompting the SMF to begin the restoration process. Instead of treating the UPF as a newly introduced network component, the SMF recognizes it as a restarted instance and proceeds with restoring its lost context. The following sections provide information on how this process ensures continuity of existing connections while effectively handling the disruption \cite{ref8}. 

\subsubsection{Extracting Values from Previous Session}

\begin{table*}[t]
    \centering
    \begin{tabular}{|c|l|c|c|l|}
        \hline
        \# & Restore Message & Sender & Receiver & Data Transferred \\
        \hline
        \multirow{3}{*}{1} & \multirow{3}{*}{\textbf{N4 Session Establishment Request}} & \multirow{3}{*}{SMF} & \multirow{3}{*}{UPF} & DNN - S-NSSAI \\
        \cline{5-5}
        & & & & Packet Detection Rules (PDR) \\
        \cline{5-5}
        & & & & Forwarding Action Rule (FAR) \\
        \hline
        2 & \textbf{N4 Session Establishment Response} & UPF & SMF & UL F-TEID \\
        \hline
        3 & \textbf{N4 Session Modification Request} & SMF & UPF & DL F-TEID \\
        \hline
        4 & \textbf{N4 Session Modification Response} & UPF & SMF & Confirmation of modification \\
        \hline
    \end{tabular}
    \caption{Restore Messages and Data Transferred}
    \label{tab:n4_messages}
\end{table*}

Following detection of a UPF failure, the SMF first examines the context of the session, a data structure maintained by the SMF that contains all the relevant information about active PDU sessions. By filtering this session context using the UPF node ID, the SMF can identify which PDU sessions were affected by the failure. It then extracts the necessary information from these sessions to support the restoration process. For better understanding, the extracted data can be grouped into three categories based on its origin:

\begin{itemize}
    \item UE Source Information: During the initial PDU session establishment in a 5G network, the UE provides the network with essential subscription and service-related information. This includes the UE's unique identifier (SUPI), the name of the data network it wants to access (DNN), and details about the network slice it wishes to use (S-NSSAI). The SMF stores this information as part of the session context, from which we extracted it.

    \item SMF Source Information: During the initial session establishment, the SMF generates and stores specific control parameters such as a unique PDU Session ID, the type of PDU session, and the Session Context Identifier (SCID).

    \item AMF Source Information: This information relates to tunnel parameters on the gNB side that are relayed to the SMF via the AMF during the session establishment process. 
\end{itemize}
    
    It includes key elements such as the QFI, which identifies and manages the quality of service for specific flows, and the DL F-TEID which is essential for establishing the downlink user plane tunnel between the gNB and the UPF.

    \subsubsection{Executing the Fault-Tolerance Procedure}
    Following the information-gathering phase, the context from the prior sessions must be transferred to the new UPF. This transfer process resembles the signaling between the SMF and UPF in the PDU session establishment procedure. Among all the signaling outlined in the PDU session establishment process, we focused on four specific signalings between the UPF and the SMF, which are based on 3GPP standards and are critical for transferring context between these two network functions. These signalings ensure seamless session restoration using data collected from previous signalings, without involving the AMF. This process involves four sequential procedures, as outlined in Table. \ref{tab:n4_messages}. 
    
    First, the SMF sends an \textbf{N4 Session Establishment Request} to the UPF to establish the uplink tunnel endpoint. This step involves creating uplink PDR(s) containing information to classify traffic of PDUs arriving at the UPF, and FARs, which contain information on whether forwarding, dropping, or buffering is to be applied to traffic identified by PDR(s) for this 5G session. In response, the UPF will generate a UL F-TEID for the session tunnel and transmit it back to the SMF. The SMF sends an \textbf{N4 Session Modification Request} to the UPF. The primary goal of this signaling is to ensure that the gNB, identified by its DL F-TEID, is recognized by the UPF as the tunnel endpoint for this PDU session. Finally, the session modification response is sent from the UPF to the SMF, indicating the session is finalized.

\subsection{Discussion}

\subsubsection{UPF Recovery Scenarios}
We have two options for UPF models: a single UPF restart and an active-standby model.

\begin{itemize}
    \item Active-Standby Scenario:
        In this model, one UPF actively handles user-plane traffic while a secondary UPF remains in standby as a backup, ready to take over seamlessly in case of failure. The recovery procedure in this scenario comprises (1) transferring the context to the backup UPF and (2) updating the context in the gNB for the tunnel, as one end entity has now changed.
        Although this approach can minimize service interruption and reduce latency, it incurs additional signaling overhead and requires enhancements in the gNB to support dynamic tunnel reconfiguration.
    \item Single UPF Restart Scenario:
        In this model, a failed UPF instance is restarted, and the context is transferred to it once it is operational again. While this method may result in longer downtime, it offers reduced complexity. 
        Ultimately, this approach was selected as it aligns more closely with our design objectives, offering a practical balance between system complexity and reliability.
        
\end{itemize}

\subsubsection{Starting Point}
The fault tolerance procedure can be initiated from multiple starting points within the session establishment process, depending on which module detects the UPF failure. Potential starting points include:
\begin{itemize}
    \item[1] UE: The procedure could start from the UE if it detects a failure, initiating a re-establishment of the session. However, this approach would introduce additional signaling and delay, as the UE would need to renegotiate the session from scratch.

    \item[2] gNB: The gNB, upon detecting an issue with the UPF, could initiate the recovery process. While this method allows for an early response, it would still require additional signaling exchanges with the control plane, adding complexity.

    \item[3] SMF: The SMF is the closest control plane function to the UPF and has access to both the previous session context and the real-time status of the UPF. By choosing the SMF as the starting point, signaling overhead would be minimized while ensuring that the UE experiences minimal disruption.
\end{itemize}

Considering these factors, we select the SMF as the optimal starting point for initiating fault tolerance, ensuring a seamless recovery process with minimal impact on ongoing user sessions.
\subsubsection{Fault Detection}
The SMF monitors the status of the UPF through two mechanisms:
\begin{itemize}
    \item[1] Heartbeat Messages: The SMF periodically exchanges heartbeat messages with the UPF to verify its availability. However, if the UPF restarts quickly, within an interval shorter than the heartbeat period, the SMF might not detect the failure.
    \item[2] Startup Notification from the UPF: When a UPF comes online, it sends a specific message to the SMF, similar to a registration request. This provides an immediate indication that the UPF has restarted.
\end{itemize}

Given the high restart speed of the UPF in our infrastructure, we rely on the second method to detect failures, ensuring that the SMF promptly recognizes when a UPF has restarted.

\subsubsection{Failure-Aware UPF Selection Mechanism}
While our failure recovery mechanism ensures service continuity for UEs already connected to the network, an additional challenge arises when selecting a UPF for newly arriving UEs. Traditional selection algorithms focus on factors such as load balancing and geographic proximity, but do not account for recent failure history. This omission can lead to users being assigned to a UPF that has recently experienced failures and may still be unstable.

To address this, we introduce a failure-aware UPF selection mechanism that considers not only the current operational status of UPFs but also their failure history. By prioritizing UPFs with a stable track record, our approach minimizes the risk of repeated failures and enhances overall network reliability. This mechanism operates in three main steps:

\begin{itemize}
    \item[1] Failure Detection: The system identifies UPF failures through status notifications. When a UPF restarts, the SMF distinguishes between a new deployment and a recovery event, then updates failure records accordingly.

    \item[2] Updating Failure Information: Failure history is maintained, tracking both the number and timing of failures. It helps to identify unstable UPFs and informs the selection process.

    \item[3] Failure-Aware Selection: When assigning a UPF to a new UE, the system prioritizes nodes with fewer recent failures. To achieve this, the list of available UPFs is sorted first by failure count (ascending) and then by time since last failure (descending). In practice, this means that UPFs with fewer failures and longer stable operation are considered more reliable and are placed higher in the selection order.

    After sorting, the system selects a UPF by ensuring compatibility with the requested slice (SNSSAI) and DNN. The chosen UPF must support both N3 (access)
     and N6 (exit) interfaces for the required configuration. By combining failure awareness with compatibility checks, this approach ensures that UEs are assigned to stable and suitable UPF nodes, enhancing overall network resilience.
\end{itemize}


\section{Experimental Results: UPF Failure and Recovery Analysis}
 This section presents our evaluation of the UPF fault tolerance and recovery mechanisms in a 5G network. Our goal is to demonstrate how the network behaves during UPF downtime and how connectivity is subsequently restored, with a focus on delay, bandwidth, overall performance and session continuity.

We analyze UPF failure and recovery using OpenAirInterface (OAI) 5G Core \footnote{https://gitlab.eurecom.fr/oai/cn5g/oai-cn5g-fed} v1.5.0 on Ubuntu 20.04 (kernel 5.15.0-122-generic). The setup integrates oai-gnb and oai-nr-ue simulators to emulate both RAN and UE, enabling end-to-end 5G evaluation. Network functions are orchestrated with Docker 27.1.1 and Docker-Compose 1.29.2.

\subsubsection{Network Architecture and Modules}

\begin{figure}[htbp]
    \centering
    \includegraphics[width=\linewidth]{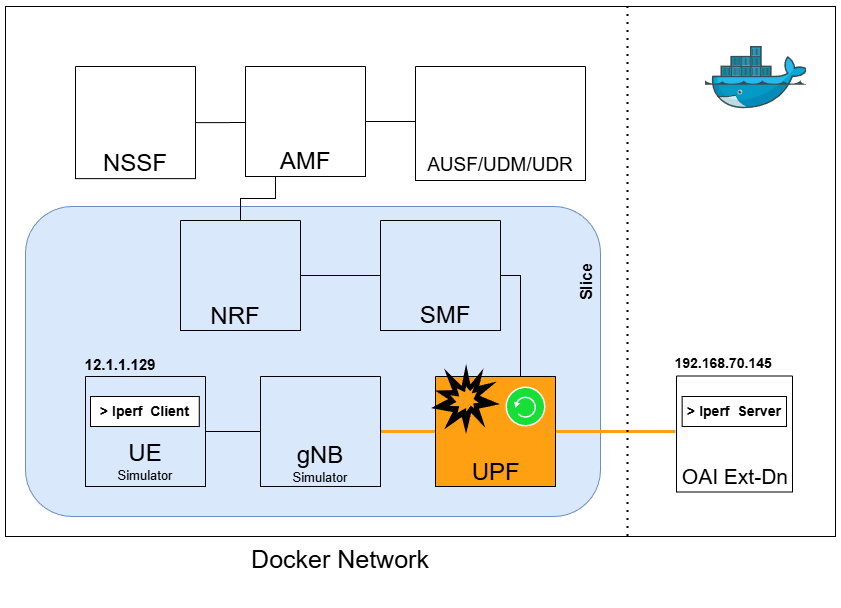} 
    \caption{Overall setup of the experiment based on the OAI 5G network slicing scenario with the entire deployment running on a single computer.}
    \label{fig:fig4}
\end{figure}

As shown in Figure~\ref{fig:fig4}, we adopted the network slicing structure provided by OAI in a fully containerized Docker environment. One of the available slices was deployed, including a dedicated SMF and a corresponding UPF based on the spgwu implementation. This slice is fully supported by the OAI gNodeB and UE simulators, enabling end-to-end validation of the 5G core functionalities. OAI simulators were used for both the UE and gNodeB components, with the UPF connected to an external data network. An additional container named Ext-Dn was included to emulate the external network and was tested during the experiments. The UPF follows the spgwu implementation with equivalent functionality.

\subsubsection{Custom SMF implementation}
To implement our session restoration mechanism, the strategy required modifications only in the SMF module. Therefore, we modified the OAI SMF source code \footnote{https://gitlab.eurecom.fr/oai/cn5g/oai-cn5g-smf/}, built a custom Docker image, and integrated it into the docker-compose setup. All other components rely on the official OAI images.

A detailed overview of the OAI codebase, highlighting the structures most relevant to our fault-tolerance implementation, is provided in Appendix~\ref{Sec5}, followed by a comprehensive explanation of our custom implementation in Appendix~\ref{Sec6}.

To evaluate the fault tolerance mechanism, we conducted experiments to measure network performance during UPF failure and subsequent recovery. Using \textit{iperf}, we tested both TCP and UDP traffic, capturing delay and bandwidth in two phases: (i) during UPF failure, which caused a temporary service interruption, and (ii) after the system restored connectivity by reselecting the original UPF and transferring the session context. In addition, we used \textit{ping} to evaluate packet loss and latency, and employed a custom monitoring script to track automatic UPF failover.

In all experiments, UPF failure was simulated by stopping and restarting the UPF container, resulting in the loss of its runtime context. This was done under the assumption that the underlying infrastructure supports automatic container restart. In the final test, however, we extended the evaluation to assess the recovery time in a scenario where a new UPF instance was instantiated by the infrastructure with the same IP address, instead of restarting the same container.

\subsection{Session Restoration after UPF Failure}

\begin{figure}[!htbp]
	\centering
	\includegraphics[width=0.46\textwidth]{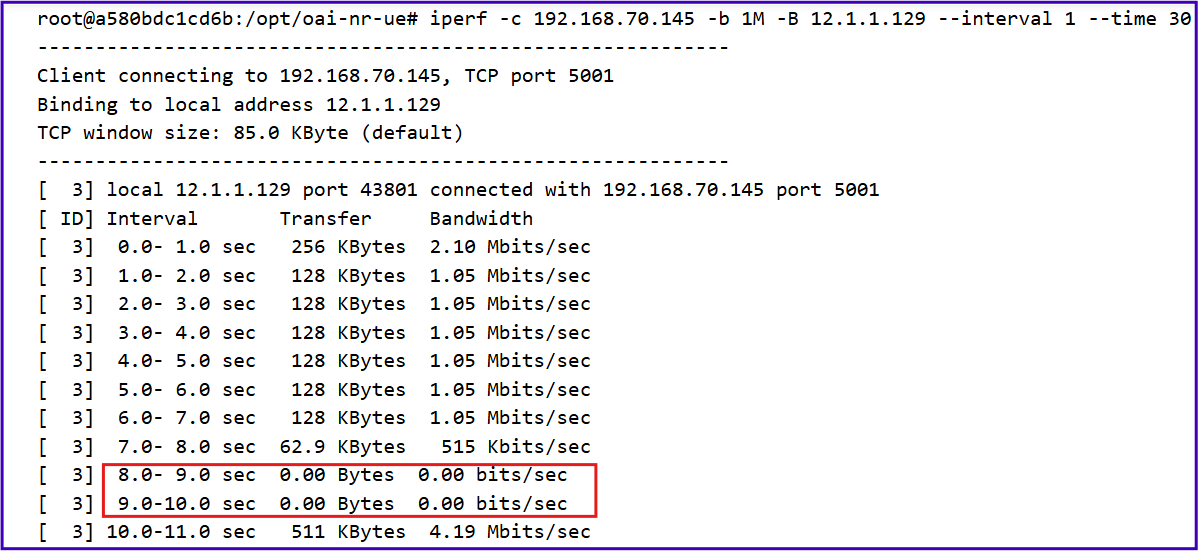}
		\caption{Performance of TCP traffic during UPF failure and recovery}
		\label{fig:fig5}
\end{figure}

In the TCP test, a failure in the active UPF resulted in a significant bandwidth drop. Initially, the TCP connection maintained a stable bandwidth of approximately 1.05 Mbps. At second 8, throughput dropped to zero, indicating a connectivity disruption. The logs in Fig. \ref{fig:fig5}  show that the interruption lasted about 2 seconds, until second 10, when the recovery mechanism was triggered.
Once the SMF restored the session with the restarted UPF, the TCP flow resumed immediately. The bandwidth increased to over 4 Mbps and subsequently stabilized around 1.05 Mbps, indicating successful recovery and re-establishment of the session with the UPF.

Server-side logs also confirmed similar observations, with an initial drop and subsequent stabilization. Overall, the results show that the recovery process brought the connection back in 2 seconds, with only a brief pause in data transfer.
\subsection{Packet Loss and Connectivity Analysis}

The UDP test was conducted like the TCP test, and the results aligned with the findings from TCP traffic. The logs for UDP traffic show a stable bandwidth of approximately 1.05 Mbps, with no significant interruptions. However, near the end of the test, a warning indicated packet loss during the UPF recovery phase:


\begin{figure}[!htbp]
        \centering{\includegraphics[width=0.46\textwidth]{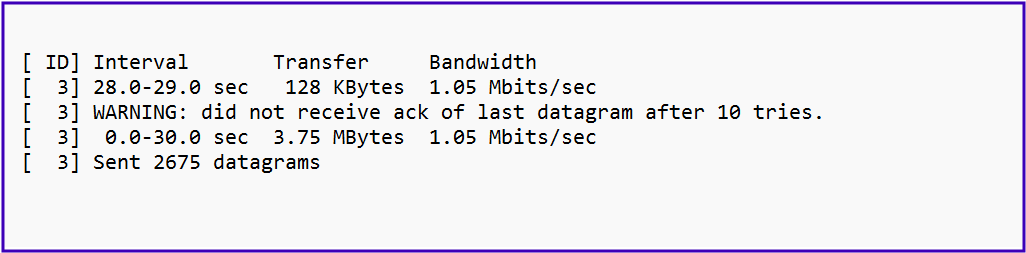}}
	\caption{Performance of UDP traffic during UPF failure and recovery\label{fig:fig6}}
\end{figure}

UDP server-side logs also confirmed continuous connectivity without major downtime. The slight packet loss observed towards the end reflects the inherent challenge of maintaining seamless UDP traffic during UPF switching. Despite the stable bandwidth, maintaining consistent UDP delivery during recovery poses challenges due to the protocol's connectionless nature.

Unlike TCP, UDP lacks mechanisms to ensure data integrity and delivery order, making it the application's responsibility to handle these aspects. UDP does not notify the sender if packets are dropped, as it does not track them. 
\subsection{Download Traffic}

To better understand the effect of a UPF failure on connectivity, we performed a ping test from the Ext-Dn module inside the UE container, which is deployed as a separate container within the network. The results confirmed that packet loss was minimal even during downtime. Out of 111 packets transmitted, 93 were received, resulting in 16.2\% packet loss, even though the UPF was briefly unavailable:


\begin{figure}[!htbp]
        \centering{\includegraphics[width=0.46\textwidth]{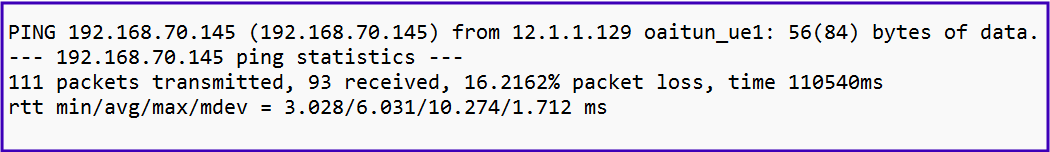}}
	\caption{Download traffic during UPF failure and recoverys\label{fig:fig7}}
\end{figure}
The RTT values provide additional insight into latency during the failure window. The minimum RTT was 3.0 ms, which is consistent with the baseline latency observed in steady-state operation. The average RTT of 6.0 ms and maximum RTT of 10.3 ms indicate only a moderate increase compared to baseline, while the standard deviation (mdev) of 1.7 ms shows that jitter remained low. Taken together, these results mean that although some packets were lost, the packets that did arrive were delivered with delays comparable to normal operating conditions, and no large spikes in latency were introduced during recovery.

This means the UPF failure did not cause a full connectivity outage but rather a partial disruption: most packets were still delivered, and those that arrived experienced stable latency. From an application point of view, that’s an important distinction: lightweight checks like ping will show missing packets, but longer TCP flows can continue after a short disruption, since TCP retransmits the lost data once the path is back.

These findings suggest that the recovery mechanism helps preserve a baseline level of connectivity, even while the UPF is being restored.

\subsection{Automated Recovery via UPF Replacement using a monitoring script}
Unlike the previous experiment, where recovery was achieved by restarting the same UPF container, this scenario was designed to test a setup with \textbf{two UPFs}. The script continuously checks the health status of the active UPF container (oai-spgwu-slice2-1). If the container is no longer reported as healthy, the script immediately starts a backup UPF container (oai-spgwu-slice2-2) but assigns it the same IP address as the failed one. This avoids the need to reconfigure other network functions, since they can continue sending traffic to the same IP. The sequence of events captured by the logs is shown in Fig.\ref{fig:fig8}:

\begin{figure}[!htbp]
        \centering{\includegraphics[width=0.46\textwidth]{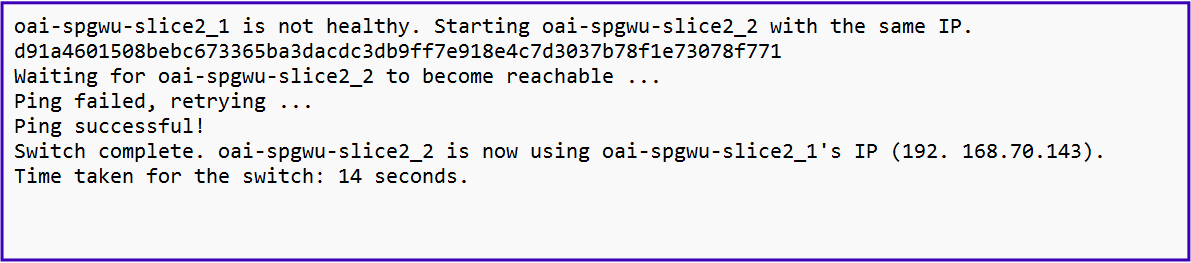}}
	\caption{Monitoring script output illustrating automated recovery by launching a new UPF container with the same IP address\label{fig:fig8}}
\end{figure}

After launching the new container, the script repeatedly pings Ext-Dn container's IP from the UE interface until connectivity is restored. The time between detecting the failure and the first successful ping is recorded as the recovery time. In our test run, this process took about 14 seconds. The duration includes container startup, IP assignment, and the time needed for the new UPF to become operational and start forwarding packets.

This approach shows that even with a health check and start-up mechanism, recovery can be automated without manual intervention. The longer recovery time compared to the previous restart-based mechanism reflects the overhead of deploying a new container and initializing it within the OAI environment. Nevertheless, the result demonstrates that container-based replacement is a feasible recovery method when built-in restart mechanisms are unavailable or insufficient.
\subsection{Summary}

Across TCP, UDP, ICMP, and automated monitoring tests, our fault tolerance mechanism effectively managed UPF failures with minimal packet loss and rapid service recovery. The TCP results displayed a brief bandwidth peak post-recovery, while UDP throughput remained stable. The ping test highlighted minimal packet loss and confirmed low latency throughout. Meanwhile, the automated monitoring and recovery script significantly reduced downtime, completing UPF recovery in just 14 seconds and maintaining the session’s continuity.

\section{Conclusion \& Future Work}\label{Sec7} 
This work introduced and evaluated a lightweight fault-tolerance mechanism for the UPF in 5G networks. By modifying the SMF to support application-layer session restoration, we showed that user sessions can be recovered after UPF failures without requiring complex redundancy schemes. In addition, we developed a monitoring script that automated UPF restart and recovery, enabling fast service restoration and demonstrating the practicality of automated recovery in real-world deployments. Since this was implemented and tested directly on the OpenAirInterface codebase, the results highlight that our approach is both practical and effective in real-world deployments.

Alongside session restoration, we also designed a failure-aware UPF selection mechanism that prioritizes nodes with fewer past failures and longer stable operation times. While the current implementation provides a foundation for smarter UPF selection, it remains an open area for further improvement. Future work could refine the ranking criteria and integrate load and latency awareness.

Unlike prior solutions that rely heavily on complex redundancy or focus primarily on the control plane, our design offers an effective yet lightweight solution for the data plane. The failure-aware UPF selection mechanism also adds resilience over time by avoiding unstable nodes, thereby improving long-term reliability.

This framework can be expanded to support an Active-Standby UPF model, where a new UPF with a different IP address replaces the failed UPF. Implementing this requires the capability for tunnel modification in the gNB to redirect packets to the new uplink endpoint. Furthermore, this approach can be extended to incorporate a UPF pool, allowing for greater flexibility within the network architecture.

\appendices

\section{OAI SMF Code base}\label{Sec5}
The section explores the OAI codebase in detail, focusing specifically on the implementation of the SMF module, and examining its various structures and classes. This comprehensive analysis will provide valuable insights into the underlying architecture and functionality of the code. By understanding these components, developers can enhance their workflow and streamline the development process. The overall codebase is large, so in this section we focus exclusively on the SMF module, which is central to our work.

\subsection{Structure}
In this section, we will first examine the overall software architecture and define the main tasks. Then, we will explore how these components interact with each other.
\subsubsection{Software Architecture}
\begin{figure}[htbp]
	\centering{\includegraphics[width=0.46\textwidth]{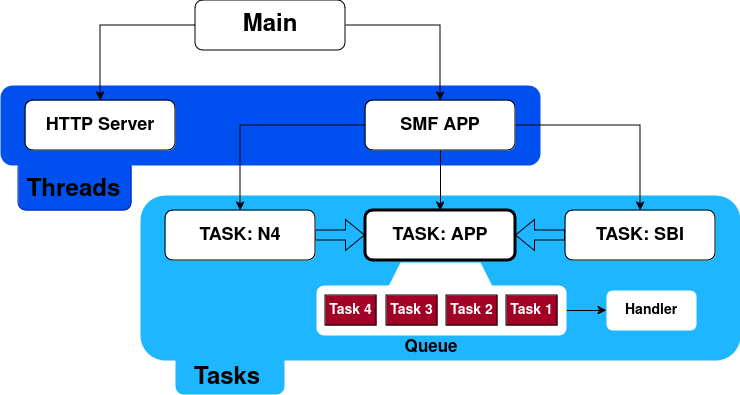}}
	\caption{SMF Primary Components\label{fig:fig15}}
\end{figure}

The primary components of the SMF as software are illustrated in Fig. \ref{fig:fig15}. Upon launching the program, two primary threads will be initiated to work:
\begin{itemize}
    \item HTTP Server: This component serves as the SBI interface for handling HTTP requests directed to other network modules. 
    \item SMF APP: The SMF Application serves as the primary component that interacts with the HTTP server, executing procedures for each request and generating appropriate responses.
\end{itemize}

After starting the threads, the SMF Application will execute three tasks that work together:
\begin{itemize}
    \item N4 TASK: The purpose of this task is to manage N4 messages that utilize the PFCP protocol instead of HTTP. It is responsible for sending and receiving messages, while any additional processing will be carried out in the SMF-App, as outlined in subsequent sections.
    \item SBI TASK: The SBI task manages both the reception and transmission of HTTP requests. This component also interacts with the HTTP server thread. SBI triggers the SMF-App to process these requests as needed. 

    \item APP TASK (SMF-App): This task serves as the central component of the SMF. Its logic can be divided into two main parts:
    \begin{itemize}
        \item Queue: An ordered collection of mini-tasks received from other tasks.
        \item Handler: A handler that processes mini-tasks according to their type and source, executing the relevant procedures and generating responses accordingly.
    \end{itemize}
\end{itemize}

\subsection{Inter-Task Communication}

In the OAI codebase, a mechanism known as the Inter-Task Interface (ITTI) is implemented to facilitate the sending of mini-tasks to the SMF-App from the other two tasks, where they are placed in the SMF-App's queue for sequential processing. Additionally, it supports messaging from the SMF-App back to those tasks, enabling a bidirectional flow of communication that allows for tasks to be processed and responses sent as needed.

In this mechanism, components interact with each other using ITTI messages. Each message contains details about the task to be performed in several key parts:
\begin{itemize}
    \item Msg Number
    \item Task Origin 
    \item Task Destination
    \item Msg Type
\end{itemize}

In the following sections, we will focus on the SMF-App task.

\subsection{Classes}
\begin{figure}[t]
	\centering{\includegraphics[width=0.22\textwidth]{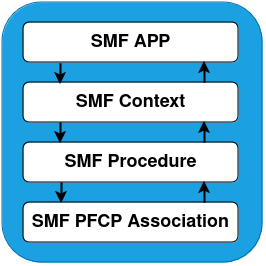}}
	\caption{SMF App Structure\label{fig:fig16}}
\end{figure}

The SMF-App can be viewed as consisting of four layers, each providing its own specific functions that interact with one another in a hierarchical manner. (see Fig. \ref{fig:fig16})

\subsubsection{SMF APP}

This layer coordinates communication among different network elements, ensuring robust session management,handling, adherence to Quality of Service (QoS) requirements, and adaptive response to network changes or failures. 
It includes two main classes:
\begin{itemize}
  \item smf\_app:
This class manages core PDU session procedures, including establishment, modification, and release, while also tracking session-related identifiers such as Session Context IDs (SCIDs). It contains specialized handlers to monitor node status, detect failures, and trigger appropriate recovery actions to maintain session integrity and minimize service disruptions.

\item smf\_context\_ref:

This class is used to manage references to specific SMF contexts. It holds identifiers like the SUPI and PDU session ID, which are essential for uniquely identifying sessions and handling procedures related to specific users and sessions.

\end{itemize}
\subsubsection{SMF Context}
This layer is responsible for handling and storing PDU sessions. It includes two main classes:
\begin{itemize}
    \item smf\_pdu\_session: This class stores the PDU session data, including DNN, SNSSAI, IP address, QoS flows, and session graph. Its functions primarily focus on getting, setting, or updating the QoS flows, which are dynamic in nature.
    \item smf\_context: This class maintains all PDU sessions associated with a specific UE and includes various handlers that perform session-related procedures in conjunction with the \textit{SMF Procedure} Class.
    
\end{itemize}
\subsubsection{SMF Procedure}
While the SMF APP and SMF Context layers oversee high-level session lifecycle management, the SMF Procedure focuses on implementing the underlying mechanisms to carry out these operations. It interacts with inter-task communication (ITTI) messages and ensures the proper application of session-related rules across network nodes.
In general, classes within the SMF Procedure layer that handle session-related tasks share a similar structure. Many of them are subclasses of either smf\_procedure or smf\_session\_procedure and implement a run function, which contains the primary logic and functionality for each specific procedure.
The SMF Procedure layer includes several key classes:
\begin{itemize}

    \item smf\_procedure: This is the base class for handling specific session procedures. It generates transaction IDs (trxn\_id) for tracking each procedure and provides the framework for responding to ITTI messages.
    
    \item smf\_session\_procedure: A subclass of smf\_procedure, this class manages session-specific procedures. It serves as a base for more specialized session management procedures.
    
    \item session\_create\_sm\_context\_procedure: Responsible for the creation of SM (Session Management) contexts, this class coordinates the establishment of PDU sessions through interactions between the SMF and external entities, like the AMF, using N4 and N11 triggers.

    \item session\_update\_sm\_context\_procedure: A subclass of smf\_session\_procedure, this class handles updates to existing session contexts. It plays a critical role in managing modifications to active sessions, including updates to QoS parameters, IP addresses, or other session-specific information that may require communication with the AMF and UPF.

    \item session\_release\_sm\_context\_procedure: This class which is also derived from smf\_session\_procedure,is dedicated to the termination and cleanup of SM contexts. It ensures that all network resources associated with a session are properly released and that notifications are sent to relevant network functions, such as the AMF, to complete the session termination.

\end{itemize}
\subsubsection{SMF PFCP Association}
This layer oversees tasks related to the UPF and consists of two primary classes:
\begin{itemize}
    \item PFCP\_Association:
    This class handles the associations between UPFs, identified by their unique node\_id, and the sessions that utilize each UPF instance as a gateway, establishing a one-to-many relationship between UPFs and sessions.
    \item UPF\_Graph:
    This class represents and manages a crucial data structure within this project. The UPF Graph is represented as an adjacency list, which is a map divided into two components, where the nodes are UPFs:
    \begin{itemize}
    \item Key: PFCP Association
    \item Value: QoS Edges:
        \begin{itemize}
        \item N3 Edge: The link between the UPF node and the gNB
        \item N6 Edge: The link between the UPF node and the Data Network
        \end{itemize}
    \end{itemize}
    This data structure will be utilized in two different ways:
    \begin{itemize}
        \item[1] PFCP Associations Class:
        This class has a single instance throughout the project and maintains the adjacency list of all UPFs in the network, along with their edge information.
        \item[2] Session Graph: This attribute belongs to the previously described pdu\_session class. Each session features a session graph that indicates which UPFs are involved in that session. Session graphs are sub-graphs of the PFCP Associations.
    \end{itemize}
    
\end{itemize}

\section{Implementation}\label{Sec6}
The primary objective is to design and implement an enhanced fault tolerance mechanism for the UPF module, which consists of two integral parts. The first part involves developing a mechanism that swiftly reacts to failure events, minimizing the impact on UE connectivity. The second part focuses on creating a system for recording detailed information about failure events, which will be utilized in future UPF selection processes. Together, these components aim to improve the overall reliability of the UPF module in handling and preventing failures. 

The implementation of the proposed fault tolerance mechanism is publicly available at:
\url{https://github.com/ION-Lab-Organization/smf}.

\subsection{Fault Tolerance Mechanism}Our mechanism was developed by leveraging existing code components as a foundation writing new functions and structures that were similar in principle but customized to our specific needs. This approach allowed us to create a tailored solution that addresses our fault tolerance requirements for the UPF module while building upon proven concepts from existing implementations. 
The pseudocode for extracting Downlink F-TEID is shown in Algorithm~\ref{alg:extract-dl-fteid}.

\begin{algorithm}[h]
\caption{Extract Downlink F-TEID}
\label{alg:extract-dl-fteid}
\begin{algorithmic}[1]
\Require $scid$ \Comment{Session Context ID}
\State Initialize a shared pointer for the PDU session: $sp \gets \varnothing$
\State Find the PDU session with $scid$ and store in $sp$
\State Retrieve the previous session graph: $prev\_graph \gets sp.get\_sessions\_graph()$
\State Get the access UPF edge from the previous graph: $access\_upf \gets prev\_graph.get\_access\_edge()$
\State Initialize an empty vector for QoS flows: $flows \gets \varnothing$
\State Get QoS flows from the access UPF edge: $access\_upf.get\_qos\_flows(flows)$
\State Retrieve the previous QoS flow: $prev\_flow \gets flows[0]$
\State Get downlink F-TEID from the previous flow: $prev\_dl\_f\_teid \gets prev\_flow.dl\_f\_teid$
\end{algorithmic}
\end{algorithm}

\subsubsection{Executing the Fault-Tolerance Procedures}
During the context transfer phase, a series of sequential network procedures must be executed. Each procedure maps to a specific run() function within the SMF procedure layer. These functions are integral to our analysis.

\begin{itemize}
    \item Session Establishment Request
    \begin{itemize}
        \item {\small \texttt{Session\_create\_sm\_context\_procedure:: \\ run}}
        \item In this function, a UPF node will be selected, and a Session Establishment Request will be sent to it. It is worth noting that the previous \texttt{node\_id} must be passed to this function to ensure the same UPF is selected for transferring the context. Following this step, the uplink tunnel context is restored in the UPF as shown in Fig. \ref{fig:fig17}.

        \begin{figure}[!htbp]
        \centering{\includegraphics[width=0.46\textwidth]{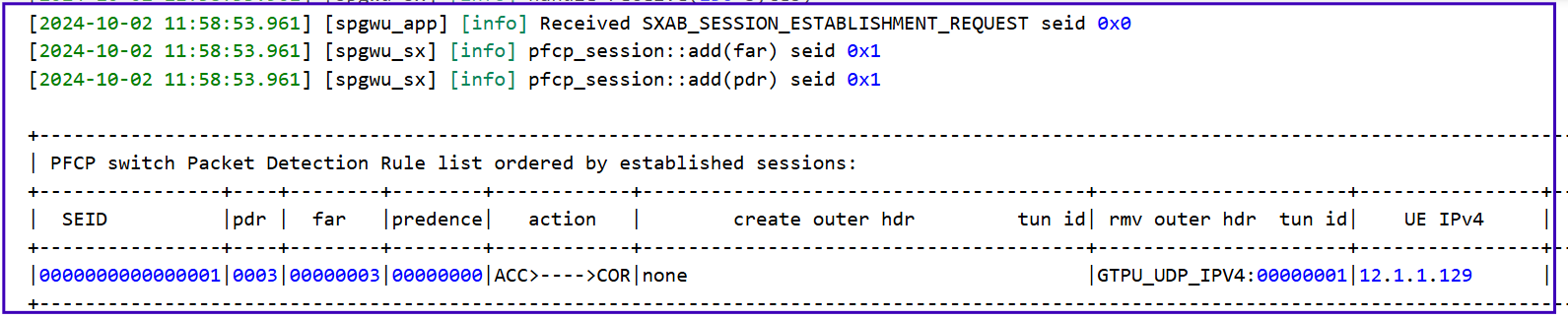}}
        	\caption{oai-spgwu logs after session establishment request\label{fig:fig17}}
    \end{figure}

    \end{itemize}

    \item Session Establishment Response
    \begin{itemize}
        \item {\small \texttt{Session\_create\_sm\_context\_procedure:: \\ handle\_itti\_msg\_restore}}
        \item Upon receiving a successful response from the UPF, the \texttt{UL\_F-TEID} and other tunnel parameters on the UPF side will be configured in the SMF.
    \end{itemize}

    \item Session Modification Request
    \begin{itemize}
        \item {\small \texttt{Session\_update\_sm\_context\_procedure:: \\ run}}
        \item In this function, the downstream tunnel information, most importantly the \texttt{DL\_F-TEID}, is sent to the UPF. While the other parameters are generated within this function, the \texttt{DL\_F-TEID} is passed along to the UPF. Following this step, the downlink tunnel context is restored in the UPF as illustrated in Fig. \ref{fig:fig18}.
        
        \begin{figure}[!htbp]
        \centering{\includegraphics[width=0.46\textwidth]{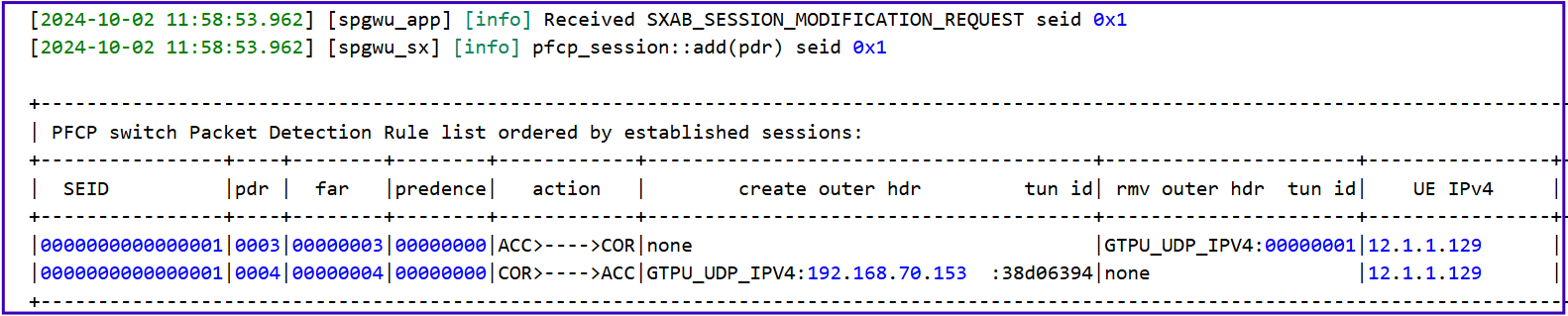}}
	\caption{oai-spgwu logs after session modification request\label{fig:fig18}}
    \end{figure}
    \end{itemize}

    \item Session Modification Response
    \begin{itemize}
        \item {\small \texttt{Session\_update\_sm\_context\_procedure:: \\ handle\_itti\_msg\_restore}}
        \item The session is finalized.
    \end{itemize}

\end{itemize}

\subsection{Failure-Aware UPF Selection Mechanism}
\subsubsection{Detecting the Failure}\mbox{}\

The \texttt{handle\_nf\_status\_notification} function is triggered by status updates from network functions, specifically when there is a change in a UPF’s operational state. This function identifies when a UPF has entered a failed state, initiating recovery actions or logging the failure for further analysis.

When a new UPF registers, the system checks if this UPF is listed in \texttt{smf\_cfg.upfs}, a vector containing registered UPF node IDs. If the UPF is already registered, indicating that it is being restarted, the system captures the \texttt{node\_id} of the failed UPF and forwards it to the \texttt{record\_failure} function.

\subsubsection{Updating Failure Information for the UPF node}\mbox{}\

The \texttt{record\_failure} function uses a custom data structure, \texttt{UpfNodeInfo}, to manage failure data for each UPF node. \texttt{UpfNodeInfo} has two key attributes:
\begin{itemize}
    \item failure\_count: This field tracks the cumulative number of failures for each UPF node. Every time \texttt{record\_failure} is called, this number increases, allowing the system to identify nodes with recurrent issues, which may cause instability.
    \item last\_failure\_time: This timestamp, recorded as a \texttt{std::chrono::time\_point}, logs the precise time of the latest failure. Capturing this time assists in analyzing failure intervals and triggering restoration if repeated failures occur within a specific timeframe.
\end{itemize}

\subsubsection{Selecting a New UPF Based on the Criteria}\mbox{}\

In the final step of UPF selection, the system assesses each available UPF node based on its failure history to ensure a stable and reliable network experience. This mechanism can be broken down as follows:
\begin{itemize}

\item{Sorting UPFs by Failure Score}

\item{Node Selection Based on Slice/DNN Compatibility:}

After sorting, the function iterates through each UPF in the sorted list to identify nodes compatible with the requested slice (SNSSAI) and DNN. For each UPF the function checks whether it supports both N3 (access) and N6 (exit), interfaces for the required slice/DNN. If both interfaces are supported, the UPF is selected as the optimal node.

\end{itemize}

\flushend
\bibliographystyle{IEEEtran}      

\bibliography{References}   
\end{document}